\documentclass[aps,prb,a4paper,twocolumn,showpacs]{revtex4}
\usepackage{graphicx,graphics,color,epsfig}
\usepackage{amsmath}
\usepackage{amsfonts}
\usepackage{amssymb}
\usepackage{slashbox}

\begin{document}
\title{Nonlocal noise cross correlation mediated by entangled Majorana
fermions}
\author{Hai-Feng L\"{u}, Hai-Zhou Lu, and Shun-Qing Shen}
\affiliation{Department of Physics, The University of Hong Kong,
Pokfulam Road, Hong Kong, China}

\begin{abstract}
Due to their nonlocality, qubits nested in Majorana bound states may
be the key to realize decoherence-free quantum computation. Majorana
bound states could be achieved at the ends of a one-dimensional
topological superconductor. However, when the bound states couple
directly to electron reservoirs their nonlocal correlation is
quenched by local Andreev reflections. Here we propose a scheme to
generate nonlocal noise cross correlation between two well-separated
quantum dots, mediated by a pair of Majorana bound states. Both
positive and negative cross correlations can be obtained by tuning
the gate voltages applied to the dots. Within a limited range of
finite temperatures, the cross correlation is not suppressed by
thermal fluctuations. Furthermore, we show how the local Andreev
reflections suppress the noise cross correlation when multiple dot
energy levels are coupled to the Majorana bound states. The
measurable cross correlation is expected to serve as a sensitive
indicator for the generation of Majorana fermions.
\end{abstract}

\date{\today}
\maketitle

\vskip2pc \narrowtext

\newpage

\section{Introduction}

Majorana fermions are exotic because they are their own antiparticles.\cite%
{s1} The search for Majorana fermions and controlling of Majorana
bound states in solid-state devices are currently attracting
increasing attention due to their potential application in quantum
information.\cite{s2,s3,s4,s5,s6,s7,s8,s9} Two well-separated
Majorana bound states can define a nonlocal fermion level and its
occupation encodes a qubit. This nonlocal topological qubit is
robust against decoherence from local perturbations or
interactions.\cite{s2,s3,s4} So far, a number of schemes have been
proposed to realize and to detect the Majorana bound
states.\cite{s10,s11,s12,s13,s14,s15,s16,s17,s18,s19,s20} For
instance, it was discovered that the Majorana bound states can be
realized at the ends of a semiconductor nanowire with strong
spin-orbit interaction in the proximity of an $s$-wave
superconductor.\cite{s10,s11,s12} Very recently, the signature for
the formation of a spatially separated pair of Majorana bound states
have been observed experimentally in such a system, in terms of a
zero-bias conductance peak.\cite{s21,s22} However, the existence of
Majorana fermions is still controversial because the zero-bias
conductance enhancement could also be the signature of diverse
phenomena in mesoscopic physics.\cite{s23,s24}

A unique feature of Majorana bound states is that their
superposition can form a nonlocal fermion level. Based on this
nonlocality, the nonlocal current-current correlation in tunneling
measurements would provide supporting evidence. Actually, the
current noise cross correlation can reveal information related to
particle fluctuations and is a powerful tool
for studying different types of interactions and quantum statistics.\cite%
{s25} For free particles, fermions tend to induce negative noise
cross correlation while bosons tend to induce positive one due to
quantum statistics of indistinguishable identical
particles.\cite{s26} In mesoscopic systems, positive cross
correlation can be induced in the presence of
Bardeen-Cooper-Schrieffer (BCS)-like
interactions,\cite{s27,s28,s29,s30} dynamical channel
blockades,\cite{s31,s32} and inelastic scattering.\cite{s33} When a
pair of Majorana bound states are coupled to two electron reservoirs
directly, positive cross correlation may be induced by the nonlocal
Andreev reflection, which means the injection of an electron into
one bound state followed by the emission of a hole by the
other.\cite{s34} However, if two Majorana fermions are well
separated, the nonlocal cross correlation might be
absent,\cite{s13,s14,s34,s35} which implies that it is hard to
obtain the nonlocal signal of Majorana fermions by electrical
measurements. This would become a serious obstacle for the
application of Majorana fermions in universal decoherence-free
quantum computing. Furthermore, unlike Cooper pairs, in which the
correlation is limited by the superconducting coherence
length,\cite{s27,s28,s29} the correlation between two Majorana
fermions is not restricted by their separation in space. In such a
real nonlocal system, nonlocal noise cross correlation is expected
to be much more important than local current autocorrelations such
as shot noise.\cite{s25} Therefore, it is of fundamental interest to
examine the noise cross correlation of two Majorana fermions in a
long distance limit.

In this paper, we propose a way to generate nonlocal current noise
cross correlation mediated by a pair of Majorana bound states.
Differing from a system where the Majorana bound states couple to
electron reservoirs directly,\cite{s13,s14,s34} two quantum dots
with single energy level are inserted between the Majorana bound
states and reservoirs. In this case, local Andreev reflections
involving a single reservoir can be suppressed by weak dot-reservoir
coupling, while crossed Andreev reflections which split a Cooper
pair over two reservoirs become dominant by strong coupling between
the Majorana bound states and quantum dots. By tuning the gate
voltages applied to the two dots, either positive or negative
nonlocal cross correlations can be induced in this hybrid system of
quantum dots and Majorana bound states. An experimental observation
of the nonlocal noise correlation can serve as an alternative proof
for the nonlocality of Majorana bound states.

The paper is organized as follows. In Sec. II, we propose the model
system for the double quantum dots and Majorana bound states and
introduce the rate equation formulism employed to study the currents
and their correlations. In Sec. III, we present the results for the
noise cross correlation between the currents flowing through the
double dots and discuss the underlying physical processes. We also
consider the effects of thermal fluctuation and local Andreev
reflections enhanced by multiple dot levels. Finally, a summary is
given in Sec. IV.

\section{Model and formulism}

\subsection{Setup}

Our proposed setup is sketched in Fig. 1. A two-dimensional electron gas at
a semiconductor heterojunction, which is intensively studied in the context
of both spin-orbit coupling and quantum dot, serves as the host for both the
Majorana bound states and the dots. In the central region, spin-orbit
coupling and strong Zeeman splitting give rise to a spinless energy band,
which resembles a $p+ip $ topological superconductor when adjacent to an $s$%
-wave superconductor wire due to the proximity
effect.\cite{s10,s11,s12} The nontrivial topological nature of the
$p+ip$ superconductor wire demands a pair of Majorana bound states
to emerge at the two ends. Two quantum dots can be fabricated by
depleting the electrons beneath the gate electrodes. Each quantum
dot is coupled to a Majorana bound state and connected with the
electron reservoir on its side. Only one spinless energy level is
considered in each quantum dot by assuming sufficiently large level
spacing and strong Zeeman field.

\begin{figure}[tbph]
\centering \includegraphics[width=0.45\textwidth]{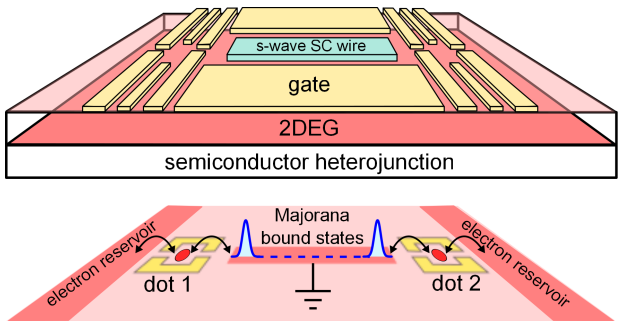}
\caption{Setup with the double quantum dots and Majorana bound states. An
\emph{s}-wave superconductor wire is deposited on the surface of a
semiconductor heterojunction, and a magnetic field is applied along the axis
of the wire. For a proper magnetic field, two Majorana bound states appear
at the ends of the topological superconductor. Two quantum dots are placed
between the bound states and the reservoirs to block the local Andreev
reflection between the Majorana bound states and electron reservoirs. The
energy level of each dot can be modulated by applying a gate voltage.}
\label{fig:setup}
\end{figure}

\subsection{Hamiltonian}

The Hamiltonian for the system in Fig. \ref{fig:setup} is given by
\begin{eqnarray}
H_0 &=&\epsilon _{1}d_{1}^{\dagger }d_{1}+\epsilon _{2}d_{2}^{\dagger }d_{2}+%
\frac{i}{2}\epsilon _{M}\eta _{1}\eta _{2}  \notag \\
&&+(\lambda _{1}^{\ast }d_{1}^{\dagger }-\lambda _{1}d_{1})\eta
_{1}+i(\lambda _{2}^{\ast }d_{2}^{\dagger }+\lambda _{2}d_{2})\eta _{2},
\end{eqnarray}%
where $\epsilon _{i}$ is the energy level in quantum dot $i$ ($i=1, 2$) and $%
d_{i}(d_i^{\dag })$ is the annihilation (creation) operator of
electron. The
quantum dots are coupled to the Majorana bound states with strength of $%
\lambda _{1}$ and $\lambda _{2}$. $\epsilon _{M}$ is the "coupling
strength" between the Majorana fermions $\eta _{1}(=\eta _{1}^{\dag
})$ and $\eta _{2}(=\eta _{2}^{\dag })$ in the Majorana bound
states. The electron reservoirs and their coupling to the dots are
described by the Hamiltonian
\begin{eqnarray}
H_{T}=\sum\limits_{ik}\epsilon _{i,k}c_{ik}^{\dagger
}c_{ik}+\sum\limits_{i,k}(t_{i}c_{ik}^{\dagger }d_{i}+h.c.),
\end{eqnarray}
where $\epsilon _{i,k}$ is the electron energy in the reservoir $i$ and $%
t_{i}$ is the tunneling amplitude.

It is helpful to replace the Majorana fermion operators by a fermion
operator, $\eta _{1}=f+f^{\dagger }$, $\eta _{2}=i(f^{\dagger }-f)$, where $%
f^{\dag }$ creates a nonlocal fermion and $f^\dagger f=0, 1$ counts
the occupation of the corresponding state.\cite{s13,s17} In the new
representation, the Hamiltonian in the central region
becomes\cite{s36}
\begin{eqnarray}
H_0 &=&\epsilon _{1}d_{1}^{\dagger }d_{1}+\epsilon _{2}d_{2}^{\dagger
}d_{2}+\epsilon_M(f^{\dagger }f-\frac{1}{2})  \notag \\
&&+[\lambda _{1}(f^{\dagger }d_{1}+fd_{1})+\lambda _{2}(f^{\dagger }d_{2}
-fd_{2})+h.c.].
\end{eqnarray}

The Hamiltonian $H_0$ can be solved in the space spanned by eight
basis states $|n_{1}n_{2}p\rangle $, where $n_{i}=0,1$ is the
electron occupation number in the quantum dot $i$. For the Majorana
bound states, the particle number in a superconducting states is not
conserved, but the parity $p$ can serve as the quantum number, with
the values of even ($e$) or odd ($o$). For the double dots and
Majorana bound states, the energy eigenvalues and eigenstates are
given in two closed subspaces of definite parity. It is
found that the states $|00e\rangle$, $|10o\rangle$, $|01o\rangle$, and $%
|11e\rangle$ form a closed even block, and other four states $|00o\rangle$%
, $|10e\rangle$, $|01e\rangle$, and $|11o\rangle$ form a closed odd
block. The eigenstates of even parity are
\begin{equation*}
|e_{l}\rangle =a_{l}^{e}|00e\rangle +b_{l}^{e}|10o\rangle
+c_{l}^{e}|01o\rangle +d_{l}^{e}|11e\rangle
\end{equation*}%
and those of odd parity are
\begin{equation*}
|o_{l}\rangle =a_{l}^{o}|00o\rangle +b_{l}^{o}|10e\rangle
+c_{l}^{o}|01e\rangle +d_{l}^{o}|11o\rangle
\end{equation*}%
with $l=1,2,3,4$ and where $a,b,c,d$ are normalized superposition
coefficients. For the two parities, their eigenequations are
\begin{eqnarray}
\mathbf{M}^{o(e)}\Psi^{o(e)}_l=E^{o(e)}_l\Psi^{o(e)}_l,
\end{eqnarray}
where
\begin{eqnarray}
\mathbf{M}^{e}=\left(
\begin{array}{cccccc}
-\frac{\epsilon_M}{2} & \lambda_1 & -\lambda_2 & 0 &  &  \\
\lambda_1^\ast & \epsilon_1+\frac{\epsilon_M}{2} & 0 & \lambda_2 &  &  \\
-\lambda_2^\ast & 0 & \epsilon_2+\frac{\epsilon_M}{2} & \lambda_1 &  &  \\
0 & \lambda_2^\ast & \lambda_1^\ast & \epsilon_1+ \epsilon_2-\frac{\epsilon_M%
}{2} &  &  \\
&  &  &  &  &
\end{array}
\right),
\end{eqnarray}
\begin{eqnarray}
\mathbf{M}^{o}=\left(
\begin{array}{cccccc}
\frac{\epsilon_M}{2} & \lambda_1 & \lambda_2 & 0 &  &  \\
\lambda_1^\ast & \epsilon_1-\frac{\epsilon_M}{2} & 0 & -\lambda_2 &  &  \\
\lambda_2^\ast & 0 & \epsilon_2-\frac{\epsilon_M}{2} & \lambda_1 &  &  \\
0 & -\lambda_2^\ast & \lambda_1^\ast & \epsilon_1+ \epsilon_2+\frac{%
\epsilon_M}{2} &  &  \\
&  &  &  &  &
\end{array}
\right),
\end{eqnarray}
and $\Psi^{o(e)}_l=(a^{o(e)}_l, b^{o(e)}_l, c^{o(e)}_l, d^{o(e)}_l)^\mathbf{T%
}$.

In the absence of the electron tunneling between the quantum dots
and reservoirs, there is no mixing between the states of different
parities. The parity in the hybrid system of Majorana bound states
and dots can be varied by one-particle sequential tunneling between
the quantum dots and the reservoirs. Thus, the even- and odd-parity
sections are, thus, connected as
\begin{eqnarray}
\langle e_l|d_1|o_{l^{\prime }}\rangle={a^e_l}^\ast b^o_{l^{\prime }}+{c^e_l}%
^\ast d^o_{l^{\prime }},  \notag \\
\langle e_l|d_1^\dagger|o_{l^{\prime }}\rangle={b^e_l}^\ast a^o_{l^{\prime
}}+{d^e_l}^\ast c^o_{l^{\prime }},
\end{eqnarray}
for the tunneling events through dot $1$ and
\begin{eqnarray}
\langle e_l|d_2|o_{l^{\prime }}\rangle={a^e_l}^\ast c^o_{l^{\prime }}+{b^e_l}%
^\ast d^o_{l^{\prime }},  \notag \\
\langle e_l|d_2^\dagger|o_{l^{\prime }}\rangle={c^e_l}^\ast a^o_{l^{\prime
}}+{d^e_l}^\ast b^o_{l^{\prime }},
\end{eqnarray}
for dot $2$. The dot-reservoir tunneling does not mix the states of the same
parity, i.e., $\langle e_l|d_{1(2)}^\dagger|e_{l^{\prime }}\rangle=0$ and $%
\langle o_l|d_{1(2)}^\dagger|o_{l^{\prime }}\rangle=0$.

\subsection{Rate equations}

We study the tunneling currents and their noise cross correlation in
the sequential tunneling regime with the help of the rate equation
method.\cite{s37} Formally, there are two types of approaches: the
generalized master equations and diagonalized rate equations. In the
generalized master equations, the density matrix contains not only
the population of each state (may not be the eigenstates) but also
the coherent terms between any two states. The diagonalized  rate
equation approach is to, first, diagonalize the Hamiltonian of
central region and then write down the rate equations for the
population of central eigenstates. For weak coupling between the
dots and reservoirs, it is safe to construct the density matrix
$\rho $ with the eigenstates of the hybrid system of Majorana bound
states and dots then write down the rate equations for the
population of the system eigenstates. It has been shown that when
the inner interaction of the central region is much stronger than
the dot-lead coupling strength, the results obtained from two
approaches are in close agreement with each other.\cite{s37,s38}
Therefore, we take $\lambda_i\gg t_i$ in the calculation to ensure
that the diagonalized rate equations work well. The time evolution
of the density matrix $\rho (t)$ is given by the rate equations
\begin{eqnarray}
d\rho (t)/dt=\mathbf{W}\rho (t),
\end{eqnarray}
where the rate matrix elements are\cite{s37,s39}
\begin{eqnarray}  \label{Wll}
W_{l^{\prime }l}=\sum_{i}\Gamma _{i}[f(\Delta _{l^{\prime }l}+\mu
_{i})|\langle l^{\prime }|d_{i}|l\rangle |^{2}  \notag \\
+f(\Delta _{l^{\prime }l}-\mu _{i})|\langle l^{\prime }|d_{i}^{\dagger
}|l\rangle |^{2}]
\end{eqnarray}
for $l\neq l^{\prime }$ and
\begin{eqnarray}
W_{ll}=-\sum_{l^{\prime }\neq l}^{N}W_{l^{\prime }l},
\end{eqnarray}
with $l,l^{\prime }\in \{|o_{l}\rangle ,|e_{l}\rangle \}$. Here
$f(\omega )=[1+e^{\omega /k_{B}T}]^{-1}$ is the Fermi distribution
function, $\Delta _{l^{\prime }l}=E_{l^{\prime }}-E_{l}$ is the Bohr
frequency of the transition from $|l\rangle $ to $|l^{\prime
}\rangle $, $E_{l}$ is the eigenenergy of $|l\rangle $, and $\mu
_{i}$ is the chemical potential in reservoir $i$. In the wide-band
limit approximation, the coupling between the $i$-th dot level and
its reservoir is denoted by $\Gamma _{i}=2\pi |t_{i}|^{2}\nu _{i}$
with $\nu _{i}$ the spinless density of states near the Fermi
surface of reservoir $i$. The current $I_{i}$ flowing through dot
$i$ is calculated by
\begin{eqnarray}
I_{i}=e\sum_{l}[\hat{\mathbf{\Gamma }}^{i}\rho ]_{l},
\end{eqnarray}
where $\hat{\mathbf{\Gamma }}^{i}$ is the matrix form of the current
operator and its elements are given by
\begin{eqnarray}
\hat{\Gamma}_{l^{\prime }l}^{i}&=&\Gamma _{i}\left[ f(\Delta _{l^{\prime
}l}+\mu _{i})|\langle l^{\prime }|d_{i}|l\rangle |^{2}\right.  \notag \\
&&\left.-f(\Delta _{l^{\prime }l}-\mu _{i})|\langle l^{\prime
}|d_{i}^{\dagger }|l\rangle |^{2}\right].
\end{eqnarray}

The noise cross correlation of two tunneling currents can be studied
by the Fourier transform of the current-current correlation
function\cite{s25}
\begin{eqnarray}
S_{12}(\omega)=2\int_{-\infty}^{\infty} dte^{i\omega t}(\langle
I_{1}(t)I_{2}(0)\rangle -\langle I_{1}\rangle\langle I_{2}\rangle ),
\end{eqnarray}
where $\left\langle \cdots \right\rangle $ represents the
thermodynamic average. Furthermore, the current-current correlation
function of the currents $I_{1}$ and $I_{2}$ in the $\omega$ space
can be expressed as
\begin{eqnarray}
\langle I_{1}(t) I_{2}(0) \rangle_{\omega} =\sum_{k} \left [\hat{\Gamma}_{1}%
\hat{T}(\omega)\hat{\Gamma}_{2}\rho^{(0)} + \hat{\Gamma}_{2}\hat{T}(-\omega)%
\hat{\Gamma}_{1}\rho^{(0)}\right ]_{k},  \notag \\
\end{eqnarray}
where $\hat{T}(\pm\omega)=\left(\mp i\omega \mathbf{I}-\mathbf{W}\right)^{-1}
$ and $\mathbf{I}$ is the unit matrix.

In the following discussion, we focus on the situation that the two
Majorana fermions are well separated so that $\epsilon _{M}=0$. The
two dots are assumed symmetrically coupled to the reservoirs
$\Gamma_i=\Gamma_0$. Considering the symmetric couplings between
Majorana bound states and dots on the opposite ends, we take
$\lambda _{i}=\lambda_{0}$.

\section{RESULTS AND DISCUSSIONs}

\subsection{Tunable nonlocal noise cross correlation}

The zero-frequency noise cross correlation $S_{12}$ is presented in
Fig. 2 as a function of dot levels $\epsilon _{1}$ and $\epsilon
_{2}$, where the voltage $V=-\mu_i$ is symmetrically applied to both
dots. The correlation shows a quadrupole pattern of four regions,
two positive and two negative. At the point $\epsilon _{1}=\epsilon
_{2}=0$, the cross correlation is zero. Away from this point, the
cross correlation with tunable signs can be induced by varying the
dot energy levels. Along the diagonal directions, the correlation
reaches the maxima. When both quantum dot levels are higher or lower
than the chemical potential (set as zero energy point) in the $p+ip$
superconductor, the correlation is negative. Otherwise, the
correlation is positive. Similar patterns were also obtained in the
capacitively coupled double dots,\cite{s32} where the cross
correlation arises from the competition between the inter- and
intradot Coulomb interactions. Here the sign reversal of the cross
correlation is mediated by the Majorana bound states. Competition
and coordination between different tunneling channels are modulated
by the occupation of Majorana bound states. As shown in Fig. 2(a),
two negative cross-correlation branches lie in the competition
region
asymmetrically and the maximum negative $S_{12}$ appears at the point $%
\epsilon_i=\mu_i$. Away from this point, electron tunneling into
(out of) the dots becomes easier when $\epsilon_i < (>) \mu_i$. This
weakens the competition between tunneling currents on the opposite
ends and results in the decrease of noise cross correlation.

\begin{widetext}

\begin{figure}[tbph]
\centering \includegraphics[width=0.9\textwidth]{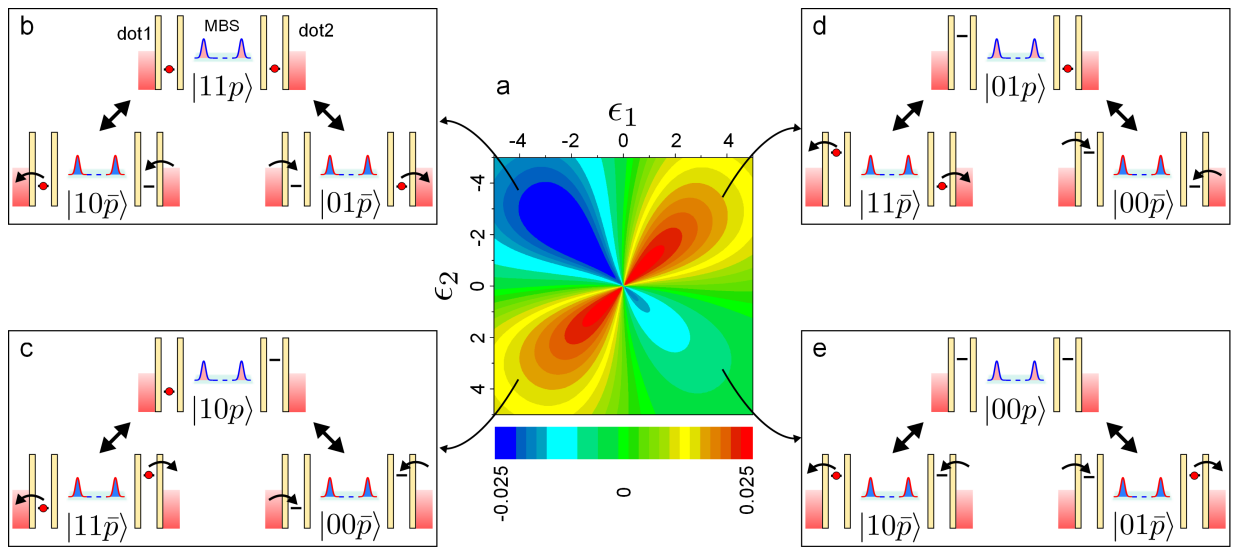}
\caption{Tunable noise cross correlation. \textbf{(a)} Zero-frequency
current noise cross correlation as a function of two dot energy
levels $\epsilon_1$ and $\epsilon_2$. The parameters are taken as
$\epsilon_M=0.0$, $k_BT=2.0$,
$\Gamma_0=0.2$, $\lambda_0=5.0$, and $\mu_i=-2.0$.
(b) When $\epsilon_1, \epsilon_2 < 0$, the system tends to reside in
the state $|n_1n_2p\rangle = |11p\rangle$,
where $n_{i}\in\{0,1\}$ is the occupation in dot $i=1,2$,
the parity $p$ of the Majorana bound states could be even or odd.
$|11p\rangle$ could evolve into either $|01\bar{p}\rangle$ or
$|10\bar{p}\rangle$, where $\bar{p}$ stands for
the opposite parity to $p$. For both $|01\bar{p}\rangle$ and
$|10\bar{p}\rangle$, electron tends to tunnel in one dot and
out of the other.
These opposite tunneling processes lead to the negative cross correlation.
Similarly, the correlation is also negative in (e) when
$\epsilon_1, \epsilon_2 > 0$.
(c) When $\epsilon_1<0<\epsilon_2$, the possible transitions are
$|10p\rangle \leftrightarrow |00\bar{p}\rangle, |11\bar{p}\rangle $,
which tend to synchronize the tunnelings in two dot-reservoir
channels and give rise to the positive correlation.
Similar processes also give the positive cross correlation in
(d) when $\epsilon_2<0<\epsilon_1$.}
\label{fig:picture}
\end{figure}

\end{widetext}

\subsection{Physical processes}

To analyze the physical processes beneath the correlation, we return to the $%
f$-representation in Eq. (3), which shows three types of transitions between
the Majorana bound states and quantum dots: pairing creation $f^{\dag
}d_{i}^{\dag }$, pairing annihilation $fd_{i}$, and direct tunneling $%
f^{\dagger }d_{i}$ and $d_{i}^{\dag }f$. We, first, take Fig. 2(b),
for example. By tuning the gate voltages, one can lower the energy
levels in both dots than the chemical potentials in the reservoirs
($\epsilon _{i}\ll \mu _{i}$). Consequently, the two dots are
occupied most of the time and the
system is in the state $|11p\rangle $ (where $p$ can be even $e$ or odd $o$%
). In this state, thermal fluctuations can induce tunneling currents
in both dot-reservoir channels, but the currents are uncorrelated
when they are not related to the Majorana bound states. For
$\epsilon _{i}\ll \mu _{i}$, the state population of $|00p \rangle$
is rather small and almost does not participate in the state
evolution. For weak dot-reservoir coupling, the states can evolute
many times before parity is changed by an electron tunneling into or
out of the reservoirs. The state $|11p\rangle $ can evolve
into $|10\bar{p}\rangle $ or $|01\bar{p} \rangle $ by direct tunneling if $%
p=e$ or pairing annihilation if $p=o$, where $\bar{p}=e$ if $p=o$
and vice
versa. In either case, the occupations in the two dots change from $%
|11\rangle $ into $|10\rangle $ or $|01\rangle $. As a result of the
interaction between the dots and Majorana bound states, electrons
tend to tunnel into one dot or out of the other. These opposite
tunneling behaviors lead to the negative cross correlation between
the currents of two dot-reservoir channels. Similarly, for the case
$\epsilon _{i}\gg \mu _{i}$ [Fig. 2(e)], the initial state
$|00p\rangle $ can evolve into $|10\bar{p}\rangle $ or $|01
\bar{p}\rangle $, which also gives rise to the negative cross
correlation. In contrast, when the zero energy point in the region
of Majorana bound states is much higher than one dot energy level
while much lower than the other, the possible transitions become
from $|10p\rangle $ to $|00\bar{p}\rangle $ or $|11\bar{p}\rangle $,
as shown in Fig. 2(c), and from $|01p\rangle $ to $|00\bar{p}\rangle
$ or $|11\bar{p} \rangle $, as shown in Fig. 2(d). These transitions
tend to synchronize the electron tunneling processes in both dots
and make the cross correlation positive.

It is indicated in Fig. 2(a) that the noise cross correlation
$S_{12}$ reaches the maximum values but with opposite signs on the
two diagonal lines $\epsilon_1=\epsilon_2$ and
$\epsilon_1=-\epsilon_2$. In these two cases, we can understand the
nonlocal correlation from analytical results. In the case
$\epsilon_1=\epsilon_2=\epsilon_0$, the eigenvalues of $H_0$ can be
found as
\begin{eqnarray}
E^{p}_{1\pm}=\frac{1}{2}\left(3\epsilon_0\pm\sqrt{\epsilon_0^2+8\lambda_0^2}%
\right),  \notag \\
E^{p}_{2\pm}=\frac{1}{2}\left(\epsilon_0\pm\sqrt{\epsilon_0^2+8\lambda_0^2}%
\right),
\end{eqnarray}
and the corresponding eigenstates are
\begin{eqnarray}
\Psi^{p}_{1\pm}&=& \varsigma|10\bar{p}\rangle+|01\bar{p}\rangle +\frac{%
\epsilon_0\pm\sqrt{\epsilon_0^2+8\lambda_0^2}}{2\lambda_0}|11\bar{p}\rangle,
\notag \\
\Psi^{p}_{2\pm}&=& \frac{\epsilon_0\mp\sqrt{ \epsilon_0^2+8\lambda_0^2}}{%
2\lambda_0}|00\bar{p}\rangle-|10\bar{p}\rangle +\varsigma|01\bar{p}\rangle,
\end{eqnarray}
where $\varsigma=1$ for even parity and $\varsigma=-1$ for odd parity. When
a relatively small voltage is applied to both dots, the eigenstate $%
\Psi^{o(e)}_{1-}$ ($\Psi^{o(e)}_{2-}$) is the ground state in the region of
Majorana bound states and dots if the dot energy levels are lower (higher)
than the chemical potential in the nanowire, i.e., $\epsilon_0<(>)0$. For $%
\epsilon_0<0$, the stationary population of $\Psi^{o(e)}_{1-}$ is
dominant and the transition between these two degenerate states
could be caused by the tunneling between dots and reservoirs. When
there is no interaction between the Majorana bound states and dots,
the ground state is the one that the each dot is always singly
occupied most of the time. The dot-Majorana interaction correlates
the two dots nonlocally and induces the transition events
$|11p\rangle\leftrightarrow |01\bar{p}\rangle$ or
$|10\bar{p}\rangle$. This means that the role of the coupling
between the Majorana bound states and dots is to turn the same state
in the two dots into opposite simultaneously. Therefore, the
tunneling events between two channels tend to repulse each other,
which gives rise to a negative noise cross correlation. Similarly,
the ground state of the system is $\Psi^{o(e)}_{2-}$ in the case of
$\epsilon_0>0$. The interaction between the Majorana
bound states and dots causes the transition events $|00p\rangle%
\leftrightarrow |01\bar{p}\rangle$ or $|10\bar{p}\rangle$. The noise
cross correlation is also negative in this case.

When the two dot levels have opposite energies $\epsilon_1=-\epsilon_2=%
\epsilon_0$, the eigenvalues of $H_0$ are
\begin{eqnarray}
E^{p}_{1\pm}=\frac{1}{2}\left(-\epsilon_0\pm\sqrt{\epsilon_0^2+8\lambda_0^2}%
\right),  \notag \\
E^{p}_{2\pm}=\frac{1}{2}\left(\epsilon_0\pm\sqrt{\epsilon_0^2+8\lambda_0^2}%
\right),
\end{eqnarray}
and the corresponding eigenstates are
\begin{eqnarray}
\Psi^{p}_{1\pm}&=&\varsigma|00\bar{p}\rangle+\frac{\epsilon_0\mp \sqrt{%
\epsilon_0^2+8\lambda_0^2}}{2\lambda_0}|01\bar{p}\rangle-|11\bar{p}\rangle,
\notag \\
\Psi^{p}_{2\pm}&=&|00\bar{p}\rangle+\frac{\epsilon_0\pm \sqrt{%
\epsilon_0^2+8\lambda_0^2}}{2\lambda_0}|10\bar{p}\rangle+\varsigma|11\bar{p}%
\rangle.
\end{eqnarray}
It is clearly shown that the coupling between the Majorana bound states and
dots turns the opposite states of double dots into the same simultaneously,
e.g., $|01p\rangle$ $(|10p\rangle)\leftrightarrow |00\bar{p}\rangle$ or $|11%
\bar{p}\rangle$. The presence of Majorana-dot coupling tends to
correlate the tunneling events in both dots in a bunching way.
Therefore, the noise cross correlation is always positive in this
case.

\subsection{Effects of thermal fluctuation}

\begin{figure}[ht]
\centerline{\epsfig{file=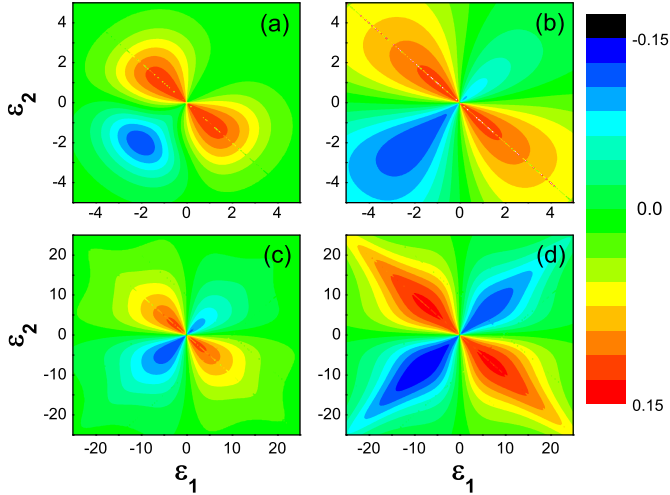,height=8.0cm} }
\caption{ The current noise cross correlation as a function of two dot
energy levels $\protect\epsilon_1$ and $\protect\epsilon_2$ for different
temperatures $k_BT$. (a) $k_BT=0.5$, (b) $k_BT=2.0$, (c) $k_BT=5.0$, and (d)
$k_BT=10.0$, The parameters are taken as $\Gamma_0=1.0$, $\protect\lambda%
_0=5.0$, and $\protect\mu_i=-2.0$. }
\label{dispers}
\end{figure}

In the sequential tunneling regime, the effect of thermal
fluctuation is reflected in the Fermi distribution of tunneling
electrons in both leads. The reservoirs are not affected by the
coupling to the dots so they remain in their respective thermal
equilibrium. With the decrease of temperature, fewer electrons away
from the Fermi energy contribute to the current and thermal
fluctuation-induced tunneling is suppressed. Figure 3 shows that the
noise cross correlation $S_{12}$ exhibits quite different patterns
for different temperatures $k_BT$. At low temperatures, the decay
time of the lead correlation is mainly determined by the bias
voltage, while the contribution from the quasiparticle lifetime is
irrelevant.\cite{s40,s41} In the calculation, we take a bias voltage
much larger than the dot-reservoir strength, which ensures that the
results at relatively low temperatures are still justified. Figure
3(a) represents the results when the temperature is lower than the
dot-lead coupling strength. In the region of negative values, the
cross correlation decreases to zero for $\epsilon_i>0$ and shrinks
to an elliptical region centering at $\epsilon_0=\mu_i$. Considered
individually, the Majorana-dot Hamiltonian $H_0$ has the
particle-hole symmetry. However, when the reservoirs with tunable
chemical potentials are connected to the dot, the particle-hole
symmetry is broken. In the calculation, we take the chemical
potential in both leads $\mu=-2.0$ to generate two currents. Here
the broken particle-hole symmetry is attributed to the finite bias
voltage and the cross correlations thus differ in the regimes of
$\epsilon_i>0$ and $\epsilon_i<0$. With the increase of temperature,
four regions in the cross-correlation pattern become wider and more
symmetric about the point $(0, 0)$. From Eq. (\ref{Wll}) it is found
that the tunneling probability of electrons is governed by the Fermi
distribution function $f(\Delta _{l^{\prime }l}\pm\mu_i)$. When
$k_BT$ is much larger than $\lambda_0$, the tunneling probability is
determined by the ratio $\epsilon_i/k_BT$. Therefore, the area of
cross correlation is proportional to the energy scale of the
temperature approximately, as shown in Fig. 3.

It is interesting that the temperature does not suppress the
strength of cross correlation, which differs from the case of
capacitively coupled double dots.\cite{s32} For interdot Coulomb
interaction-induced cross correlations in double dots, each dot is
connected to two electron reservoirs. At a relatively high
temperature, the thermal fluctuation helps the electrons tunnel into
or out of the dots more easily. The (anti-)blockade effect between
two tunneling channels is, thus, weakened, leading to a suppression
of noise cross correlation. However, for the present device, one of
the reservoirs is replaced by one end of the topological
superconductor. The zero-mode Majorana fermions serve as the source
or drain, in which only one fermionic level takes part in the
transport. Although the temperature determines the state population
of the system, the mechanism of Majorana mediated cross correlation
is independent of the temperature.

\subsection{Suppression of nonlocal noise correlation by enhanced local
Andreev reflection via multiple dot levels}

In the setup where the Majorana bound states couple directly to the
reservoirs, the noise cross correlation disappears ($S_{12}=0$) for
vanishing $\epsilon _{M}$.\cite{s13,s14,s34} Strictly speaking,
$\epsilon_{M} $ must be $0$ according to the definition that the
Majorana fermions are their own antiparticles. In the present device
the direct couplings between the Majorana bound states and
reservoirs are blocked by the quantum dots. The weak dot-reservoir
coupling suppresses the local Andreev reflection between the
reservoirs and Majorana bound states  and enhances the nonlocal
cross correlation. To test this picture, we also study two cases in
which multiple dot energy levels are coupled to the Majorana bound
states: (1) only one end of the topological superconductor is
coupled to multiple levels and; (2) both ends are coupled to
multiple levels. The two cases are schematically shown in Figs. 4(a)
and 4(b), respectively. The energy levels provide more conducting
channels between the reservoirs and Majorana bound states, thus
enhancing the local Andreev reflection. As the number $N$ of the
energy levels approaches infinity, the model becomes equivalent to
the case when the Majorana bound states are coupled to electron
reservoirs directly. As a result, the current noise cross
correlation is expected to vanish.

\begin{figure}[tbph]
\centering \includegraphics[width=0.4\textwidth]{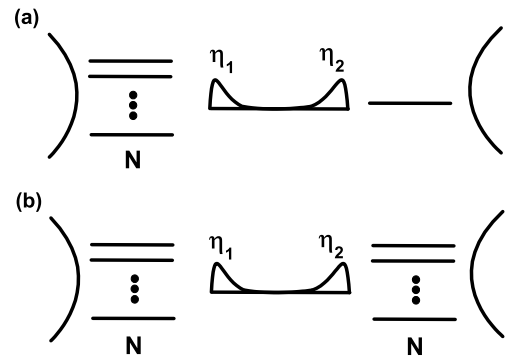}
\caption{Schematic view for the cases when multiple levels are coupled to
the Majorana bound states at the ends of the topological superconductor.
Multiple levels are coupled to (a) only one end of the topological
superconductor and (b) both ends. Each level is assumed to have the same
energy and coupling to the Majorana bound states.}
\label{fig:picture3}
\end{figure}

\begin{figure}[tbph]
\centering \includegraphics[width=0.5\textwidth]{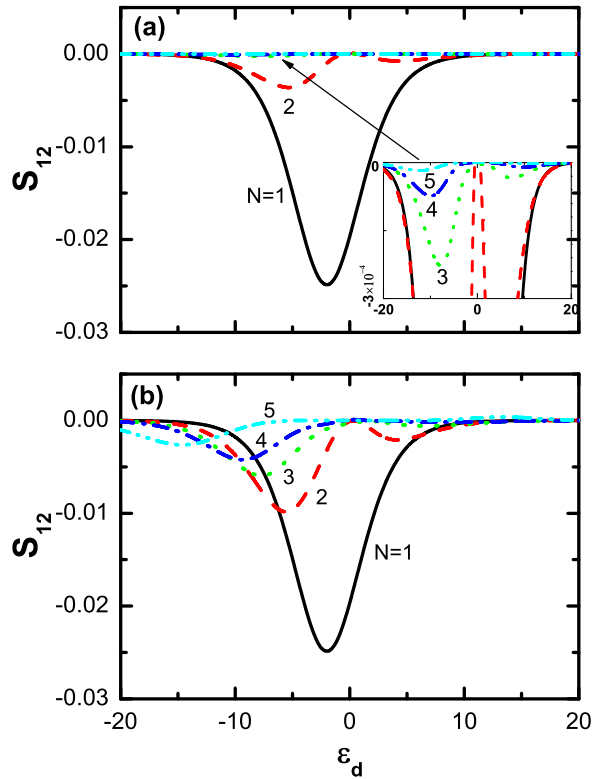}
\caption{Current noise cross correlation as a function of energy level $%
\protect\epsilon_d$ in the presence of multiple levels, where $N$ counts the
level number. The $N$ levels couple to only one end (a) and both ends (b) of
the topological superconductor, where the coupling strength between the $N$ levels and
reservoirs is set as $\Gamma=\Gamma_0/N$. The parameters are taken as $%
\protect\epsilon_M=0.0$, $k_BT=2.0$, $\Gamma_0=0.2$, $\protect\lambda_0=10.0$%
, and $\protect\mu_i=-2.0$.}
\label{fig:picture4}
\end{figure}

Figure 5 demonstrates the noise cross correlation $S_{12}$ as a
function of the energy level $\epsilon_d$ for different level
numbers $N$. For simplicity, it is assumed that the levels are of
the same energy and coupling strength to the Majorana bound states.
For comparison, Fig. 5(a) presents the result for the first case in
Fig. 4(a), in which only one end couples to multiple levels, while
Fig. 5(b) gives the result of the second case in Fig. 4(b) where
both ends couple to multiple levels. For $N=1$, the model reduces to
the case along $\epsilon_1=\epsilon_2$ in Fig. 2. In this case, the
maximum negative $S_{12}$ can be achieved. It is found that with
increasing number of energy levels, the cross correlation $S_{12}$
decreases rapidly in both cases. In the presence of multiple levels,
the electron tunneling caused by Andreev reflection does not have to
always involve both electron reservoirs  but can occur via two
channels in the same reservoir. In this way, the cross correlation
between the currents flowing through two reservoirs is suppressed.
Since the Andreev reflection can occur via any two levels, it is
more likely to occur at the end of more energy levels for the case
in Fig. 4(a). As a result, the suppression of $S_{12}$ by extra
levels in Fig. 5(a) is much stronger than that in Fig. 5(b) as the
level number $N$ increases.

\section{summary}

We propose a setup to generate nonlocal noise cross correlation
between the currents flowing via two well-separated quantum dots,
mediated by a pair of entangled Majorana fermions. Each dot is
inserted between a Majorana bound state and a nearby electron
reservoir to suppress the local Andreev reflection between them. In
this case, the crossed Andreev reflection over two dots becomes
dominant due to the strong coupling between the Majorana bound
states and dots. It is demonstrated that the nonlocal cross
correlation with tunable signs can be induced by the Majorana-dot
coupling and modulated by the dot energy levels. Because the
nonlocal correlation cannot be induced by other mechanisms, these
measurable noise cross correlations can serve as a unique signature
for the formation of Majorana bound states. It is found that the
cross correlation is not suppressed with the increase of temperature
and the area of correlation is proportional to the temperature
approximately. Moreover, we study the case when multiple dot energy
levels are coupled to the Majorana bound states. It is shown that
with the increase of energy level number, the local Andreev
reflection becomes enhanced and suppresses the crossed Andreev
reflection considerably. As a result, the current noise cross
correlation vanishes gradually.

We thank T. K. Ng and K. T. Law for insightful discussions. This
project was supported by the Research Grant Council under Grant No.
HKUST3/CRF/09, the Foundation for Innovative Research Groups of the
NSFC under Grant No. 61021061, and the NSFC (Grants No. 11004022 and
No. 61006081).

\end{document}